\documentclass[smallextended]{svjour3}       
\smartqed  
\usepackage{graphicx}
\usepackage{epstopdf}
\begin{document}

\title{Dynamic nuclear polarization and relaxation of H and D atoms in solid
mixtures of hydrogen isotopes}

\author{S. Sheludiakov$^{1}$, J. Ahokas$^{1}$, J. J\"arvinen$^{1}$, O. Vainio$^{1}$,
L. Lehtonen$^{1}$, S. Vasiliev$^{1}$, D.M.
Lee$^{2}$ and V.V. Khmelenko$^{2}$}
\date{\today}

\institute{$^{1}$Wihuri Physical Laboratory, Department of Physics and Astronomy,
University of Turku, 20014 Turku, Finland\\
$^{2}$Institute for Quantum Science and Engineering, Department of
Physics and Astronomy, Texas A\&M University, College Station, TX,
77843, USA}
\maketitle
\begin{abstract}
We report on a study of Dynamic Nuclear Polarization and electron
and nuclear spin relaxation of atomic hydrogen and deuterium in solid molecular matrices of H$_{2}$, D$_{2}$, and HD mixtures. The electron
and nuclear spin relaxation times ($T_{1e}$ and $T_{1N}$) were measured
within the temperature range 0.15-2.5$\,$K in a magnetic field of 4.6 T, conditions which ensure a high polarization of electron spins. We found that $T_{1e}$ is nearly temperature independent in this temperature range, while $T_{1N}$ decreased
by 2 orders of magnitude. Such strong temperature dependence is typical for the nuclear Orbach mechanism of relaxation via the electron spins. We found that the nuclear spins of H atoms
in solid D$_{2}$ and D$_{2}:$HD can be efficiently polarized by
the Overhauser effect. Pumping the forbidden transitions of H atoms also leads to DNP, with the efficiency strongly dependent on the concentration of D atoms. This behaviour indicates the Cross effect mechanism of the DNP and nuclear relaxation, which turns out to be well resolved in the conditions of our experiments. Efficient DNP of H atoms was also observed when pumping the middle D line located in center of the ESR spectrum. This phenomenon can be explained in terms of clusters or pairs of H atoms with strong exchange interaction. These clusters have partially allowed transitions in the center of the ESR spectrum and DNP may be created via the resolved Cross effect. 
\end{abstract}

\section{Introduction}

Hydrogen atoms in a matrix of molecular hydrogen at
high enough concentrations may lead to a number of fascinating quantum
phenomena. Quantum correlations between atoms may lead to a BEC-like behavior of H atoms or to a transition to the conducting state of a
normally insulating H$_{2}$ matrix. Such a metal-insulator transition occurs in semiconductors, e.g. P-doped silicon at a critical concentration of [P]=$3\times10^{18}$cm$^{-3}$. Similar effects for H atoms may be expected at much higher concentrations because of the smaller size of  H atoms compared to phosphorus donors.
In our recent study of Dynamic Nuclear Polarization (DNP) of H atoms
in solid D$_{2}$ with a small admixture of H$_{2}$ and HD \cite{DNPPRL14} we observed effects, which may indicate the possible onset of the insulator-to-metal transition.  We found
that pumping exactly in the center of the ESR spectrum, where no DNP effects are expected for isolated atoms, leads to an effective polarization of the nuclear spins of H atoms. This
resembles the well-known Overhauser effect in metals where nuclear spins
become polarized after ESR pumping at the Larmor frequency of conduction
electrons. The nuclear spins relax through the hyperfine coupling
to the conduction electrons and the spin polarization emerges due
to a difference in the rates of the flip-flop  and flip-flip 
cross-relaxation \cite{Overhauser}. Similar effects were also recently
observed in insulating glassy solids doped by radicals \cite{Can2014}
and Si:P where ESR pumping of the exchange-coupled donor line led
to polarizing $^{29}$Si nuclei \cite{Dementyev_2014}. 

In this work we deal with three main methods for polarizing nuclear spins dynamically,
i.e excluding a brute force technique: the Overhauser effect (OE),
the Solid effect (SE) and the Cross effect (CE). In all these methods the allowed or forbidden transitions are induced by applied rf power, which leads to a change of the populations of the electron and nuclear spin states. As a result, the electron spin polarization is effectively  transferred to the nuclei, and therefore the DNP is most effective at low temperatures and high magnetic fields, when high degree of electron spin polarization is obtained in equilibrium.  
The Solid effect relies on saturating the forbidden transitions followed by the thermal relaxation of electron spins, and therefore requires much higher excitation power. The
Cross effect and thermal mixing are the three-spin phenomena. They
require two electron spins being coupled together by the dipolar interaction
and a nuclear spin having hyperfine or dipolar coupling to one of the electron
spins. The Cross effect may be realized
when the ESR frequencies of two electrons $\omega_{e1}$ and $\omega_{e2}$ differ by the Larmor frequency $\omega_{N}$ of a nucleus coupled to one of them \cite{Maly08}, so that: 
\begin{equation}
\omega_{e1}=\omega_{e2}+\omega_{N}\label{eq:cross}
\end{equation}
In this case saturating first electron transition at $\omega_{e1}$ leads to emergence of strong oscillating internal fields, and hence to an increase of the photon density at this frequency. This stimulates simultaneous flips of the second electron and nuclear spin, resulting in DNP. Here, unlike the Overhauser and Solid effects, pumping and cross-relaxation occur at the same frequency. Although the Cross effect is a well established method, it has been
exclusively observed for inhomogeneously broadened ESR lines having
a breadth greater than the nuclear Larmor frequency \cite{Atsarkin2012}. In this situation it cannot be resolved from the Solid effect, and the analysis of the DNP dynamics becomes rather complicated. 

Polarizing nuclear spins by the Solid effect requires much larger
oscillating fields than in the case of the Overhauser effect. Saturating
a flip-flop forbidden transition requires mm-wave fields of the order
of $(A/B)^{-2}$ larger than for the allowed transition, where $A$
is the hyperfine constant (507$\,$G for H) and $B$ is magnetic field.
The flip-flip transition is completely forbidden for the free atoms
but becomes allowed in solids by the factor $(B_{dip}/B)^{2}$ where
$B_{dip}$ is the dipolar term of the electron-nuclear interaction.
The probability of polarizing nuclear spins by the well-resolved Cross
effect for the case when requirement (\ref{eq:cross}) is fulfilled
can be substantially larger than that
for the Solid effect because the allowed ESR transition is saturated in this case \cite{Wollan76}.

In order to create large enough nuclear polarization by means of
DNP, two main requirements should be fulfilled. The electron polarization
should be close to unity and the nuclear relaxation time should be
longer than the relaxation time through the forbidden transitions.
These conditions are fulfilled in high magnetic fields
(above 1$\,$T) and low temperatures (below 1$\,$K). A comprehensive
study of the spin-lattice relaxation in Si:P, a system similar to
H in H$_{2}$, was carried out in works of Feher \cite{Feher_Gere} and
Honig \cite{Honig_Stupp} while a thorough theoretical analysis of
the mechanisms of electronic and nuclear relaxation is presented in
the book of Abragam and Goldman \cite{Abragam_Goldman}. 

In this work we will focus on clarification of mechanisms responsible
for the formation of DNP of hydrogen atoms recently observed by our
group in a D$_{2}$:H$_{2}$ matrix \cite{DNPPRL14}. We present a detailed study of the spin-lattice relaxation
and extended our experiments on DNP of H and D atoms to several other matrices, including p-H$_{2}$, o-D$_{2}$, HD and D$_{2}$:2\%HD mixtures. The
measurements are performed at the conditions of high magnetic field $B$=4.6$\,$T and
low temperatures $T\sim$100$\,$mK where electron spins are fully
polarized, while the H nuclear spin polarization is relatively low $p=(n_a - n_b)/(n_a + n_b)\sim$0.1 in thermal equilibrium. 

\begin{figure}
\includegraphics[scale=0.5]{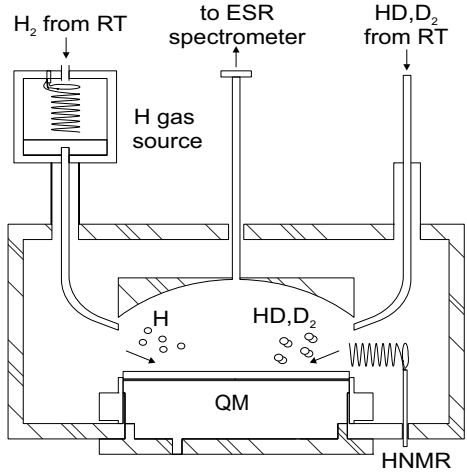}
\caption{The sample cell schematic. Here QM is the quartz crystal microbalance,
HNMR is the helical resonator used for running discharge in the sample
cell. \label{fig:The-sample-cell}}
\end{figure}

We observed a much faster relaxation of the H nuclear spins in a solid
D$_{2}$ matrix compared to other hydrogen matrices, H$_{2}$ and
HD. It turned out that the nuclear relaxation times below about 0.7$\,$K
for all cases were almost temperature independent. At higher temperatures we observed a strong temperature dependence, which may be explained by the nuclear Orbach process. We found
that nuclear polarization of H can be created in D$_{2}$, HD and
D$_{2}:$HD matrices by pumping forbidden transitions of H atoms and
the center of the ESR spectrum which match the position of D lines
in our magnetic field. An extensive study of H and D DNP in different
hydrogen matrices makes it possible for us to conclude that the presence
of unpaired D atoms becomes essential for effective polarization and
cross-relaxation of H nuclear spins. A well resolved Cross effect between H and D atoms is suggested as a possible explanation of such behaviour.

\section{Experimental setup}

The experiments were conducted in the experimental cell shown in Fig.\ref{fig:The-sample-cell}. The cell is anchored to the mixing chamber of the Oxford 2000 dilution
refrigerator and is located at the center of a 4.6$\,$T superconducting
magnet. The main diagnostic tools in our experiments are a 128$\,$GHz
ESR spectrometer and a quartz-crystal microbalance used for measuring the thickness of the hydrogen films \cite{Cellpaper}. The films of solid hydrogen isotopes are
deposited directly from the room temperature gas handling system onto
the top electrode of the quartz-crystal microbalance which also plays
a role as the bottom mirror of the Fabry-Perot ESR resonator. The temperature of the quartz surface is kept below 1 K in the deposition process, which allows growing smooth and homogeneous films \cite{Cellpaper}. 

\begin{figure}
\includegraphics[width=1\columnwidth]{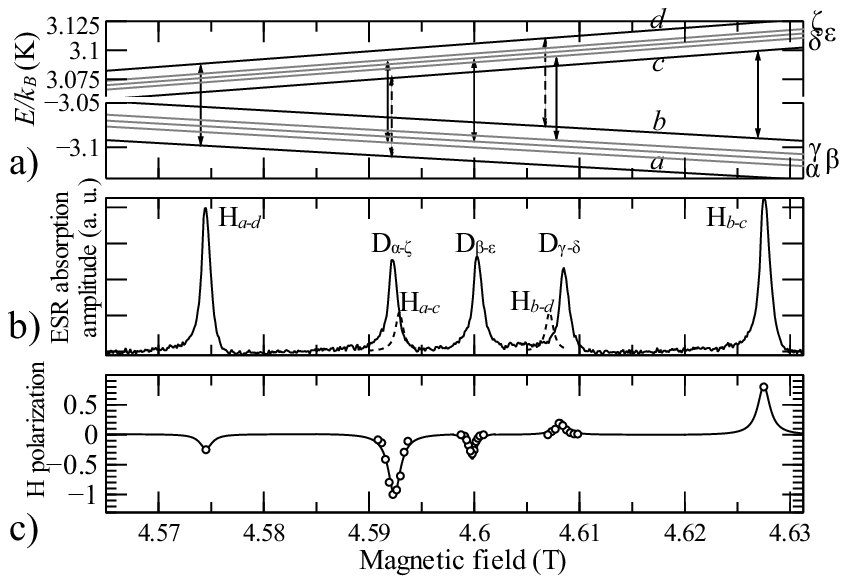}

\protect\caption{a) Energy levels and ESR transitions for atomic H an D in magnetic
field B=4.6$\,$T, b) ESR spectrum of H and D, the H forbidden transitions are shown dotted, c) Nuclear polarization of H atoms as a function of the ESR pumping position. The equilibrium polarization is subtracted for convenience. The solid line is drawn as a guide for the eye.
\label{fig:a)-Energy-levels}}
\end{figure}

Prior to depositing the samples, 
a small amount of He was condensed into
the volume below the QM in order to have a saturated film which flushes
the substrate and helps to provide a proper cooling to the samples.
After the hydrogen films are deposited, a rf discharge is started
in the sample cell using a helical resonator (HNMR) and the accumulation
of unpaired atoms begins. The films we studied were $\sim250\,$nm
thick, which is $\sim 2.5$ times larger than the penetration depth of the electrons
($\sim100\,$eV) generated during discharge. The discharge was turned
off after running it for $\sim$2 days. 
The concentrations of unpaired atoms in the films
were evaluated from the density dependent shifts and widths of the ESR lines \cite{HinH2Rapid09}. The total concentrations of H and D atoms in the samples we studied were $\sim3\times10^{19}$cm$^{-3}$ while the H:D ratio varied from sample to sample. The maximum D concentration reached 
in HD films was $\sim 3\times 10^{18}$ cm$^{-3}$. In these samples the D atom concentrations decayed with a characteristic time of $\sim3$ $\,$h after stopping the discharge due to the reaction of isotopic exchange with HD molecules \cite{Exchange_reactions}. In addition to the atoms stabilized in solid hydrogen films we were able to accumulate substantial densities of H and D atoms in the gas phase in the bulk of the sample cell. The ESR lines of these free atoms served as good field markers for measurements of spectral characteristics of the trapped atoms. This was especially useful for finding exact location of forbidden transitions of H atoms.
\section{Experimental results}
\emph{Relaxation measurements}. In order to clarify the influence of the environment on the efficiency of DNP and relaxation of impurity atoms, the measurements were performed with several distinct hydrogen matrices:
para-H$_{2}$, ortho-D$_{2}$, HD and D$_{2}$:2\%HD mixtures. We
studied both relaxation behavior as a function of temperature and
the DNP effects by pumping allowed and forbidden ESR transitions. In all matrices except pure D$_{2}$, concentrations of D atoms were much smaller than H due to the fast conversion of D atoms to H caused by the isotopic exchange reactions D+H$_{2}\rightarrow$HD+H and D+HD$\rightarrow$D$_{2}$+H. In pure HD matrices the signal of D disappeared within about $\sim3$ $\,$h after turning off the rf discharge. Therefore, the relaxation behavior
for D atoms was studied only in the D$_{2}$ sample where a large
fraction of D atoms remained unconverted into H.

The electron spin-lattice relaxation times were measured by a saturation-recovery method: first we saturated the ESR line using mm-wave power of $\approx 1 \mu$W and then measured the
line recovery after the ESR excitation was decreased by 30$\,$dB to a low enough level to neglect saturation effects. The $T_{1e}$ times for both H and D were $\sim$0.1$\,$s for all
samples we studied. The relaxation times decreased only by a factor
of 2 after raising temperature from 0.15 to 1.5$\,$K. The measurement of the nuclear relaxation was performed after substantial nuclear polarization was created by one of the DNP methods, as described in the next section. We found that the H and D nuclear relaxation times decreased significantly at
temperatures above $\sim$0.7-1$\,$K. The nuclear relaxation was faster for H in a D$_{2}$ matrix compared to para-H$_{2}$ and to the relaxation rate of D in the same D$_{2}$ sample. These
results are summarized in Fig.\ref{fig:Arrhenius}.

The measurements of the cross-relaxation were done using Overhauser DNP, by saturating the allowed $b-c$ or $a-d$ ESR transitions. The rate of the DNP growth is determined by the cross-relaxation, which was monitored as the evolution of the ESR lines as a function of the pumping time. The cross-relaxation through the flip-flop transition for H atoms turned out to be most efficient in the D$_{2}$ sample, $T_{ca}$=10$\,$s, which is significantly smaller than for H in H$_{2}$: $T_{ca}$ $\sim$10$^{3}$s. The most peculiar behavior
was observed in HD where the cross-relaxation times were dependent
on the age of the sample: they were shorter right after stopping discharge
in the sample cell and increased upon sample storage. The cross-relaxation
time through the flip-flip transition, T$_{db}$, was found to be
about 10$^{3}$s for H in D$_{2}$ but was too long to measure in
other samples. We did not observe any
dependence of the cross-relaxation times on temperature.

\emph{Dynamic nuclear polarization}. DNP for H and D atoms was performed by stopping magnetic field sweep at the value corresponding to a certain allowed or forbidden transition, and maximum excitation power of $\approx 1 \mu$W was turned on for 0.5-1 hour. This power was sufficient for partial saturation of the allowed ESR transitions. On the other hand, forbidden transitions were very far from saturation. The lines of atoms in the matrices containing deuterium were broadened by a mixture of homogeneous and inhomogeneous broadening effects. Therefore, in order to saturate the whole line, the frequency of the excitation source was modulated with a deviation of several MHz. The lines of H in H$_2$ were homogeneously broadened, and no modulation was needed to saturate the whole line by staying at its peak. 

The only way to create DNP in the para-H$_{2}$ samples was the Overhauser effect by saturating
the $b-c$ transition. We achieved H polarization
$p=$0.8 by pumping the high field line for $\approx$1$\,$h. 
We were unable to polarize H nuclear spins in para-H$_{2}$ by pumping at the positions of any other transitions of H or D. The situation was completely different for H atoms in matrices containing deuterium. Efficient DNP was observed by pumping all allowed and forbidden transitions of H and allowed transitions of D atoms. The effect did not work the other way round: by pumping allowed transitions of H we did not observe any DNP effects on D. The most efficient way to create DNP of H was to pump 
its $b-c$ transition. The $b$-state atoms were rapidly transferred into the $a$-state, and polarizations close to $p\sim0.9$ were achieved within a few minutes. Pumping the low field line worked substantially slower; $p=-0.2$ was reached after one hour of pumping. In both these cases the DNP is obtained via the OE of H. Then we performed OE DNP for D atoms. The DNP effect was also observed in this case, but its efficiency was substantially smaller than that for H atoms. In addition we found that pumping the D lines leads to effective polarization of H atoms. This was not surprising for the first and third lines, since they coincide with the locations of the H forbidden transitions, and the SE was expected to work for creating DNP of H. A completely unexpected observation was that pumping the middle line of D also leads to a negative polarization of H. We note that there are no either allowed or forbidden transitions for unpaired H atoms at this location. 

\begin{figure}
\includegraphics[width=1\columnwidth]{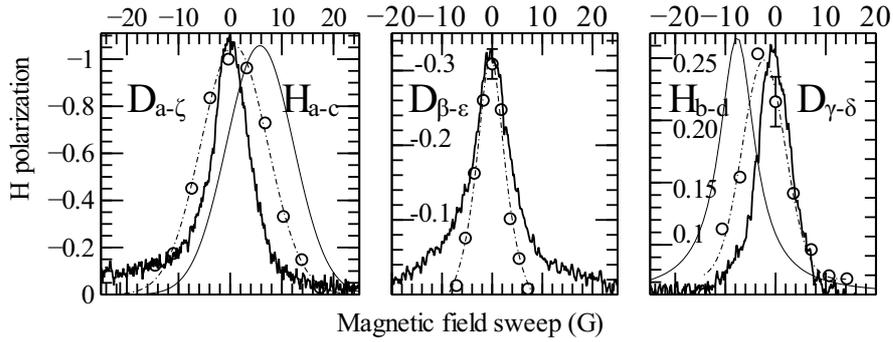}

\protect\caption{ The patterns of H atom DNP after ESR pumping near the lines of atomic deuterium in solid D$_{2}$ (open circles). The pumping times near D$_{\alpha-\zeta}$, D$_{\beta-\epsilon}$ and D$_{\gamma-\delta}$ transitions were $1.7\,$min, $10\,$min and $1$h, respectively. The positions of the H forbidden transitions are shown as a reference. The equilibrium polarization is subtracted for convenience.\label{fig:DNP-all-D}}
\end{figure}

Next we performed the measurement of the H DNP enhancement as a function of the pumping positions at the locations of all three lines of atomic deuterium for the D$_{2}$ sample. In our previous work the H DNP transition near D$_{\alpha-\zeta}$ was very broad and appeared saturated \cite{DNPPRL14}. Now we reduced the pumping time and width in order to reveal the lineshapes in more detail. The results are presented in Fig.\ref{fig:a)-Energy-levels}$\,$c and \ref{fig:DNP-all-D}. As was described earlier, pumping D$_{\alpha-\zeta}$ line and D$_{\beta-\epsilon}$ leads to a large negative polarization of H nuclear spins, while pumping D$_{\gamma-\delta}$ creates a clear positive polarization. The H polarization enhancement coincides well with the shapes of the D ESR transitions (Fig.\ref{fig:DNP-all-D}). The H DNP transition near D$_{\gamma-\delta}$ appears to be slightly shifted towards the  H$_{b-d}$ transition.

A similar behavior was observed for DNP in HD samples where the effects were somewhat weaker. The strength of these effects were also dependent on the age
of the sample; they slowly disappeared with the characteristic time of 3-4 days.
We also tried DNP with D atoms in D$_2$ samples. The only way of polarizing D spins was the Overhauser effect when the allowed transitions were pumped. No polarization enhancement appeared
after pumping forbidden transitions of D atoms.

\section{Discussion}

Magnetic relaxation at low temperatures is stimulated by the lattice phonons, and its rate
depends on the number of phonons which are in resonance, or "on speaking terms"
with the transitions between the energy levels in the spin system. The electron spins at helium temperatures relax through the direct process which corresponds to the emission or absorption
of a single resonant phonon \cite{Feher_Gere,Abragam_Goldman}.
The relaxation appears either as a spontaneous process at the lowest
temperatures when its rate is temperature independent or acquires
a linear dependence at higher temperatures which comes from the shape
of the phonon spectrum. The electron spin-lattice relaxation times
were measured previously in several experiments. It turned out that
the relaxation times measured at 0.3$\,$T \cite{Collins93,Bernard04,Dmitriev2015},
0.86$\,$T \cite{Sharnoff_Pound}, 4.6$\,$T \cite{Katunin82} and the present work
agree with each other within 1-2 orders of magnitude. Based on the assumption
that the electron-spin relaxation at helium temperatures and below
proceeds via the direct process, a strong, $\frac{1}{T_{1e}}\sim B^{4}$,
dependence is expected \cite{Honig_Stupp,Abragam_Goldman}. Although, the decrease of the relaxation time in higher fields is basically seen in the previous results, it is difficult to figure out the actual dependence due to the relatively large scatter of the data in the works cited above. However, the weak temperature dependence reported there and observed in our work provides evidence for the direct process as the main relaxation mechanism of electron spins.

The direct process for the nuclear spins at the same temperatures
is a factor of $(\omega_{e}/\omega_{N})^{3}$ less efficient than
for the case of the electron spins. This should lead to a factor of
$\sim$10$^{6}$ longer relaxation times for the nuclear spins compared
to electronic relaxation. As a result, the nuclear relaxation is driven
by the nuclear Orbach process. In this process the phonons of the lattice stimulate transitions to the upper electronic states followed by the relaxation via the forbidden transitions. The rate of this process is defined by the rate of the forbidden relaxation multiplied by the Boltzmann factor to take into account lower population of the upper states in equilibrium: 
\begin{equation}
\frac{1}{T_{N}}=(\frac{1}{T_{ca}}+\frac{1}{T_{bd}})\times exp\,(\frac{-2\mu_{B}B}{k_{B}T})\label{eq:nuclear Orbach}
\end{equation}
The electron Zeeman energy in the exponent provides a very strong temperature dependence, especially at low temperatures and strong magnetic fields, which is easily observed in our experiments.

\begin{figure}
\includegraphics{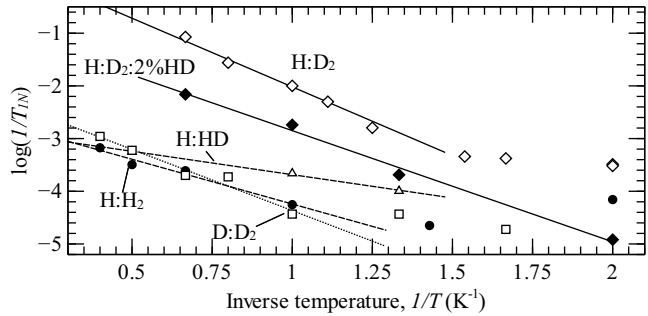}

\protect\caption{Relaxation rates plotted in Arrhenius coordinates for $T>$0.5$\,$K.
Note almost parallel slopes for H:D$_{2}$, H:D$_{2}$:2\%HD and D:D$_{2}$
($E_{a}$ $\simeq$5.5$\,$K) and different slopes for H in H$_{2}$
and H in HD\label{fig:Arrhenius}}
\end{figure}

In order to check the contribution of the nuclear Orbach process we
plotted the relaxation rates in Arrhenius coordinates (Fig.\ref{fig:Arrhenius}).
The data points for H:D$_{2}$, H:D$_{2}:$2\%HD and D:D$_{2}$ at temperatures $>$0.7 K follow almost parallel lines, which correspond to the activation energy $E_{a}$ $\simeq$5.5$\,$K,
close to the electron Zeeman energy splitting in our field $E_{Zeeman}$=6.3$\,$K. At lower temperatures, the data tend to deviate from these exponential functions, and have weaker temperature dependence. This may indicate a possible contribution from other relaxation mechanisms, e.g. the direct process. 

Comparing the nuclear relaxation and DNP efficiency for H atoms in different matrices we note a general feature observed in this work: \textit{the efficiency of these processes is greatly enhanced by the presence of D atoms in the same matrix}. Thus the nuclear relaxation rate of H in D$_2$ is two orders of magnitude faster than in pure H$_2$. At the same time, we observed that the forbidden cross-relaxation rate is also faster by two orders of magnitude for the former case. These two observations are in a fair agreement with the nuclear Orbach process, when the relaxation rates are related by equation (\ref{eq:nuclear Orbach}). We believe that such influence of D atoms on the DNP and relaxation is caused by the coincidence of allowed ESR transitions energies for D with the forbidden transitions for H atoms. This match of energies ensures an easy exchange between the H and D electron Zeeman systems \cite{Wollan76}, and stimulates cross-relaxation of H atoms. Similar behavior was observed in hydrated copper salts where the nuclear spin-lattice relaxation accelerated upon matching the condition for
the Cross effect (\ref{eq:cross}) \cite{vanHouten77_2}. 
In our case a simultaneous flip of the electron and nuclear spin of an H atom may
be accompanied by a D atom electron spin flip. This may proceed adiabatically for both forbidden transitions of H atoms because both of them have significant overlap with the deuterium lines.
However this effect is expected to be weaker for $b-d$ cross-relaxation
because the probability of this flip-flip transition is much smaller
than that for the $a-c$ transition.

The explanation of the D atom influence is also supported by our observation that the rate of the Overhauser DNP and nuclear relaxation of H in HD is decreased slowly upon
storage of the sample. This is caused by the decrease of the D atom density due to the reaction of isotopic exchange \cite{Exchange_reactions}. Although the ESR signal of D atoms vanishes in noise after 3-4 hours, substantial amounts of them are still present in the matrix, continuously decaying in time and leading to the slow decrease of the cross-relaxation rates observed in our experiments.

In Fig.\ref{fig:DNP-all-D} we presented the H DNP enhancement data together with the ESR spectra of D and calculated positions of the H forbidden transitions. The data and the ESR lines are scaled to the same amplitude for a better comparison. This picture demonstrates how closely the D lines match the locations of the H forbidden transitions in our field. One can see that the centers of the fits to the data are slightly shifted towards the forbidden transitions, where the largest overlap for these transitions occurs. The formation of H DNP
after pumping D$_{\alpha-\zeta}$ and D$_{\gamma-\delta}$ lines
may be caused by two competing effects: by the well-resolved Solid
or Cross effects. The Solid effect is a two-spin effect and involves
only an electron and a nuclear spin stimulated by a single rf photon. It does not involve any other atoms, and is not dependent on the D density. In contrast, in the Cross effect the allowed D-atom transition is saturated at the same time. Tilting the dipolar moments of D atoms leads to emergence of the oscillating magnetic fields at the exactly same frequency as required to stimulate the H forbidden transition. The density of resonant rf photons is strongly increased, which leads to an enhancement of the forbidden transitions of H. The efficiency of the CE is proportional to the D density, as is observed in our experiments. This provides evidence that the well-resolved Cross effect is the main DNP mechanism in the samples with mixed H and D atoms. To our knowledge, in all previous experiments with the Cross effect this phenomenon was studied when the allowed and forbidden transitions were not resolved, and all three DNP mechanisms OE, SE and CE occurred simultaneously. In the experiments of this work, we are able to resolve these effects, and observed the Cross effect to be well resolved from other DNP mechanisms.

However the Cross effect cannot be used to explain the DNP of H atoms
after pumping the center of the ESR spectrum (middle D-line, see Fig. \ref{fig:DNP-all-D}). There are no allowed or forbidden transitions for unpaired H atoms in this region of magnetic field. In our previous work \cite{DNPPRL14} we suggested that the
DNP observed after pumping the center of the ESR spectra
is related to the formation of radical pairs or clusters of H or H and D atoms which have strong exchange interaction between their electrons.  As is known for metals, the exchange effects lead to the averaging of the hyperfine interaction and make allowed electron spin transitions exactly at the center of the ESR spectrum, as would be the case for a free electron spin.  Overhauser DNP in metals is performed by pumping these transitions, and leads to the negative polarization of nuclei. A weak and broad ESR line from clusters and pairs of atoms is also observed for $^{31}$P donors in silicon \cite{Feher55} at sufficiently high density of donors, but still in the dielectric state. In our case the ESR signal from the pairs is too weak to be seen directly in the ESR spectrum. But the presence of D atoms via the resolved Cross effect allows their indirect detection via the DNP created by pumping the center of the ESR spectrum. As in the case of metals, negative polarization is obtained because the flip-flop transition probability is usually higher than the flip-flip probability. We believe that further studies of the DNP phenomena in high density samples of H and D atoms stabilized in solid molecular matrices will provide more evidence for these intriguing effects. 

\begin{acknowledgements}
We acknowledge the funding from the Wihuri Foundation and the Academy of Finland grants No. 258074, 260531 and 268745. This work is also supported
by NSF grant No DMR 1209255. S.S. thanks UTUGS for support.
\end{acknowledgements}

\bibliographystyle{spphys}
\bibliography{Sheludiakov_H_relaxation_JLTP_submission}

\end{document}